\documentclass[11pt]{article}

\usepackage[T1]{fontenc}
\usepackage[sc]{mathpazo}
\usepackage{amsmath}
\usepackage{amssymb}
\usepackage{enumerate}
\usepackage{amsthm}
\usepackage{amsfonts,mathrsfs}
\usepackage[style]{fncychap}
\usepackage{graphicx} 
\usepackage{geometry} 
\usepackage{eepic}
\usepackage{ifthen}
\newboolean{ElectronicVersion}
\setboolean{ElectronicVersion}{true} 

\usepackage{hyperref}
\hypersetup{pdfpagemode=UseNone}

\def\be{\begin{equation}}
\def\ee{\end{equation}}
\def\bea{\begin{eqnarray*}}
\def\eea{\end{eqnarray*}}

\def\I{\mathbb{1}}

\newenvironment{mylist}[1]{\begin{list}{}{
    \setlength{\leftmargin}{#1}
    \setlength{\rightmargin}{0mm}
    \setlength{\labelsep}{2mm}
    \setlength{\labelwidth}{8mm}
    \setlength{\itemsep}{0mm}}}
    {\end{list}}

\def\ot{\otimes}

\newcommand{\iinner}[2]{\langle #1 | #2\rangle}
\newcommand{\out}[2]{| #1\rangle\langle #2 |}

\newcommand{\pa}[1]{(#1)}

\newcommand{\ket}[1]{|#1\rangle}

\DeclareMathOperator{\trace}{Tr}
\newcommand{\ptr}[2]{\trace_{#1}\pa{#2}}

\newcommand{\tr}[1]{\ptr{}{#1}}

\newcommand{\id}{\mathbb{1}}

\def\cH{\mathcal{H}}

\def\D{\textsf{D}}

\newtheorem{thrm}{Theorem}[section]

\theoremstyle{definition}

\newtheorem{remark}[thrm]{Remark}
\newtheorem{exam}[thrm]{Example}
\numberwithin{equation}{section}

\newcounter{questionnumber}

\begin{document}

\title{\Large Mixed Form of Ambiguous and Unambiguous Discriminations}

\author{Sunho Kim,\,\, Junde Wu \\{\small \it Department of Mathematics, Zhejiang
University, Hangzhou 310027, PR~China}\\ Minhyung Cho\\{\small \it
Department of Applied Mathematics, Kumoh National Institute of
Technology, Kyungbuk, 730-701, Korea}}

\date{}
\maketitle \mbox{}\hrule\mbox\\
\begin{abstract} In this paper, we introduce a mixed form of ambiguous and unambiguous quantum state discriminations, and show that the mixed form has higher success probability than the unambiguous quantum state discriminations.
\end{abstract}
\mbox{}\hrule\mbox\\



{\bf The corresponding author}: Minhyung Cho, Department of Applied
Mathematics, Kumoh National Institute of Technology, Kyungbuk,
730-701, Korea, E-mail: chominhyung@126.com

\section{Quantum state discrimination}

Let $\cH$ be a finite dimensional complex Hilbert space. A quantum state $\rho$ of some quantum system, described
by $\cH$, is a positive semi-definite operator of trace one, in
particular, for each unit vector $\ket{\psi} \in \cH$, the
operator $\rho = \out{\psi}{\psi}$ is said to be a \emph{pure
state}. We can identify the pure state  $\out{\psi}{\psi}$ with the
unit vector $\ket{\psi}$. The set of all quantum states on $\cH$
is denoted by $\D(\cH)$.

A quantum measurement on the quantum system $\cH$ is a family of
operators $\{M_{x}\}_{x\in \Gamma}$ which are indexed by some
classical labels $x$ corresponding to the classical outcomes of the
measurement. These operators satisfy (\cite{Davies, Holevo, Kraus}):
\begin{eqnarray*}
\forall x : M_{x} \geq 0,\ \ \sum_{x} M_{x} = \I,
\end{eqnarray*}
together with $\{A_{x}\}$ such that $M_x = A_{x}^{\dag}A_{x}$.
Given a quantum state $\rho$ and a quantum measurement $\{M_x\}$, a probability distributive $p = (p_x)$ and a conditional state $\rho_{A|x}$ given outcome $x$ are induced as following:
\begin{eqnarray*}
\rho_{A|x} = p_x^{-1}A_{x}\rho A_{x}^{\dag} \ \ , \ \ p_x = Tr
(M_{x}\rho).
\end{eqnarray*}

The carriers of information in quantum communication and quantum computing are quantum systems, the information is encoded in a set of states on those systems. After processing the information, Alice transmitting it to receiver Bob. Bob has to determine the output state of the system by performing quantum measurements. If given states $\{\rho_i\}_{i\in\Sigma}$ with orthogonal supports, then it is easy to devise a quantum measurement that discriminates them without any error. However, if the states $\{\rho_i\}_{i\in\Sigma}$ are non-orthogonal, then a perfect discrimination is impossible. It is important to find the best quantum measurement to distinguish the non-orthogonal states with the smallest possible error.

Now, ones have two way for discriminating non-orthogonal states, if
the number $|\Gamma|$ of possible outcomes for quantum measurement
$\{M_{x}\}_{x\in \Gamma}$ is equal to the number $|\Sigma|$ of
states in the discriminating states, then it is called the ambiguous
quantum measurement. If $|\Gamma| = |\Sigma| + 1$ and ones can
identify perfectly each state $\rho_i$ for $|\Sigma|$ measurement
outcomes, but, there is a measurement outcome leads to an
inconclusive result (\cite{Spehner}), then it is called the
unambiguous quantum measurement.

Henceforth, for ambiguous quantum measurement, we identify the
measurement outcome with the corresponding state, thus, the outcomes
set $\Gamma$ is $\Sigma$, for unambiguous quantum measurement, we
identify the measurement outcome with the corresponding state, thus,
the outcome set $\Gamma$ is $\Sigma\cup \{0\}$, that is, for
unambiguous quantum measurement, if the outcome is $i\in\Sigma$,
then Bob is certain that the state is $\rho_i$, whereas if the
outcome is $0$, then he cannot decide what it is. Therefore, if
$\{M_{i}\}_{i\in \Sigma\cup \{0\}}$ is an unambiguous quantum
measurement, then for any $i,j\in\Sigma$, $\tr{M_i\rho_i} > 0$ and
when $i\neq j$, $\tr{M_j\rho_i} = 0$.

Let us consider an ensemble $\{\rho_i, p_i\}_{i\in\Sigma}$ of states
$\{\rho_i\}_{i\in\Sigma}$ with prior probability distribution
$p=(p_i)$. Then for each ambiguous quantum measurement
$M=\{M_i\}_{i\in\Sigma}$, the success probability of all quantum
states $\{\rho_i\}_{i\in \Sigma}$ can be discriminated is
(\cite{Spehner}) \bea P^{amb}_{suc} = \sum_{i\in\Sigma}
p_i\tr{M_i\rho_i}. \eea

For each unambiguous quantum measurement
$M=\{M_i\}_{i\in\Sigma\cup \{0\}}$, the success probability of all
quantum states $\{\rho_i\}_{i\in \Sigma}$ can be discriminated is
\bea P^{una}_{suc} = \sum_{i\in\Sigma} p_i
\tr{M_i\rho_i} = 1 - \sum_{i\in\Sigma} p_i \tr{M_{0}\rho_i}. \eea

If the probability  $p_{0} = \sum_i p_i \tr{M_{0}\rho_i}$ of
occurrence of the inconclusive outcome is minimized, then the
quantum measurement is said to be an optimal measurement.

\begin{exam}(\textbf{RRA scheme}, \cite{Roa}) Let $\cH_1 = \mathbb{C}^2$, $\{\ket{0}, \ket{1}\}$ be its orthogonal basis, $\ket{\pm} = (\ket{0} \pm
\ket{1})/\sqrt{2}$. Consider two non-orthogonal quantum states
$\ket{\psi_+}, \ket{\psi_-}\in\cH_1$ are randomly prepared with a priori probability distributive $p=(p_+,p_-)$. In order
to discriminate the two states $\ket{\psi_+}$, $\ket{\psi_-}$, taking an
auxiliary qubit system $\cH_A$, two complex numbers $c_+, c_-$ with $\overline{c_+}c_- = \iinner{\psi_-}{\psi_+}$, and prepare a quantum state
$\ket{k_a}$ in $\cH_A$, $\{\ket{0_a}, \ket{1_a}\}$ is an orthonormal
basis of $\cH_A$,  then couple $\cH_1$ to $\cH_A$ by a joint unitary
transformation $U_1$: \\
\parbox{15cm}{
\begin{eqnarray*}
U_1\ket{\psi_+}\ket{k_a} &=& \sqrt{1-|c_+|^2}\ket{+}\ket{0_a} + c_+\ket{0}\ket{1_a},\\
U_1\ket{\psi_-}\ket{k_a} &=& \sqrt{1-|c_-|^2}\ket{-}\ket{0_a} + c_-\ket{0}\ket{1_a}.
\end{eqnarray*}}\hfill
\parbox{1cm}{\begin{eqnarray}\label{eq:transformation-1}\end{eqnarray}}\\
After the joint transformation, the quantum state we consider in
discriminating is given by \be\label{eq:mixed state-1} \rho_1 =
\sum_{i=+,-}p_iU_1(\out{\psi_i}{\psi_i}\ot\out{k_a}{k_a})U_1^{\dagger}.
\ee

Note that if we perform a von Neumann measurement $\{\out{0_a}{0_a},
\out{1_a}{1_a}\}$ on the auxiliary system, then the quantum state $\rho_1$ will collapse to either $\out{0_a}{0_a}$ or
$\out{1_a}{1_a}$. If the system collapses to $\out{0_a}{0_a}$, we
will discriminate successfully the original state since we can
distinguish deterministically the two orthogonal states $\ket{\pm}$ in
(\ref{eq:transformation-1}). However, we fail if the system
collapses to $\out{1_a}{1_a}$. Thus, we can design a unambiguous quantum measurement $\prod_1 = \{\pi_i\}_{i=+,-,\textrm{0}}$ on the quantum system $\cH_1\ot\cH_A$ as follows:
\bea
\pi_+ = \out{+}{+}\ot\out{0_a}{0_a}, \quad \pi_- = \out{-}{-}\ot\out{0_a}{0_a}
\quad \textrm{and}\quad
\pi_{\textmd{0}} = \id_{\cH_1}\ot\out{1_a}{1_a},
\eea
it will unambiguous discriminate the quantum states $\ket{\psi_+}\ket{k_a}$ and $\ket{\psi_-}\ket{k_a}$, therefore $\ket{\psi_+}$ and $\ket{\psi_-}$ are unambiguous discriminated, too.

\end{exam}

The RRA scheme is extended to the case with three non-orthogonal
states in $\mathbb{C}^3$, that is:

\begin{exam}(\cite{Xu})
Let $\cH_2 = \mathbb{C}^3$, $\{\ket{0}, \ket{1}, \ket{2}\}$ be its orthogonal
basis. Ones randomly prepared three nonorthogonal states $\{\ket{u_i}: i=0, 1, 2\}$
with a priori probability distributive $p=(p_i)$, and these states
satisfy that $\iinner{u_i}{u_{j\neq i}} = \gamma_{ij}.$ In order to
discriminate the three states $\{\ket{u_i}: i=0,1,2\}$, we prepare $\{\ket{\phi_i}:
i=0,1,2\}\subseteq \cH_2$, and taking complex numbers $\alpha_i$, $\alpha_j$ such that
$\overline{\alpha_i}\alpha_j\iinner{\phi_i}{\phi_{j}} =
\gamma_{ij}$, then we couple the
original system $\cH_2$ to $\cH_A$ by the following joint unitary
transformation $U_2$:
\be\label{eq:transformation-2}
U_2\ket{u_i}\ket{k_a} = \sqrt{1-|\alpha_i|^2}\ket{i}\ket{0_a} + \alpha_i\ket{\phi_i}\ket{1_a},
\ee
where $i=0,1,2$.

If we perform the von Neumann measurement
\bea
\pi_0 = \out{0}{0}\ot\out{0_a}{0_a}, & & \pi_1 = \out{1}{1}\ot\out{0_a}{0_a}, \quad \pi_2 = \out{2}{2}\ot\out{0_a}{0_a}\\
\textrm{and} & & \pi_{\textmd{0}} = \id_{\cH_2}\ot\out{1_a}{1_a}
\eea
on the quantum system $\cH_2\ot\cH_A$, then those three states $\{\ket{u_i}\}_{i=0,1,2}$ can be unambiguous discriminated.

Now, we assume $p_2\geq p_1\geq p_0$, and let $$\gamma_1 =
\sqrt{p_1}/(\sqrt{p_2} - \sqrt{p_1}),$$ $$\gamma_2 =
\sqrt{p_0}/(\sqrt{p_2} - \sqrt{p_1}).$$

In (\cite{Xu}), the authors showed that if $\iinner{\psi_i}{\psi_{j\neq i}} = \gamma_{ij} = \gamma$, then the
maximal success probabilities of unambiguous discrimination are:

\,\ (1). If $\gamma_2\geq1$, then $P_{suc,max}^{una} = 1 - \gamma$,

\,\ (2). If $\gamma\geq\gamma_1$, then $P_{suc,max}^{una} = 1 - p_0
- p_1 - 2p_2\gamma^2/(\gamma+1)$,

\,\ (3). If $\gamma_1 \geq \gamma \geq \gamma_2$, then
$P_{suc,max}^{una} = 1 - p_0 - 2\sqrt{p_1p_2}\gamma - (\sqrt{p_2} -
\sqrt{p_1})^2\gamma^2$,

\,\ (4). If $1 \geq \gamma_2 \geq \gamma$, then $P_{suc,max}^{una} = 1 -
2(\sqrt{p_1p_2} + \sqrt{p_0p_2} - \sqrt{p_0p_1})\gamma$.
\end{exam}

In this paper, for three quantum states discrimination, we introduce a mixed form of ambiguous and unambiguous quantum state discriminations, and show that the mixed form has higher success probability than the unambiguous quantum state discriminations.

\section{Mixed form of ambiguous and unambiguous discriminations}
Firstly, we consider a special case, that is, let $\cH_2 = \mathbb{C}^3$ and prepare three states $\{\ket{u_i}\}_{i=0,1,2}$ in $\cH_2$ with a priori probability distribution $p = (p_i)$. We assume that $\iinner{u_2}{u_0} = \iinner{u_2}{u_1} = \gamma\neq0$, $\iinner{u_0}{u_1} = 0$,  where $\gamma$ is a real number. In order to  discriminate the three states $\{\ket{u_i}\}$, we define
\bea
\ket{v_i} \equiv \ket{u_i}\ket{k_a}, i = 0, 1, 2.
\eea
Taking two states $\ket{\psi_0},\ket{\psi_1}$ satisfying $\iinner{v_2}{\psi_0} = \iinner{v_2}{\psi_1} = 0$ and
\bea
\ket{v_0} &=& \sqrt{1-\gamma^2}\ket{\psi_0} + \gamma\ket{v_2},\\
\ket{v_1} &=& \sqrt{1-\gamma^2}\ket{\psi_1} + \gamma\ket{v_2}.
\eea
It follows from $\iinner{u_0}{u_1} = 0$ that $\iinner{v_0}{v_1} = 0 = (1-\gamma^2)\iinner{\psi_0}{\psi_1} + \gamma^2$. We denote \bea
c^2 \equiv \iinner{\psi_0}{\psi_1} = -\frac{\gamma^2}{1-\gamma^2}.
\eea

Similarly to the RRA scheme, we couple the original system $\cH_2$ to the auxiliary system $\cH_A$ by a joint unitary transformation $U_3$ such that $U_3\ket{v_2} = \ket{2}\ket{0_a}$ and
\bea
U_3\ket{\psi_0} &=& \sqrt{1-|c|^2}\ket{+}\ket{0_a} + \overline{c}\ket{1}\ket{1_a},\\
U_3\ket{\psi_1} &=& \sqrt{1-|c|^2}\ket{-}\ket{0_a} + c\ket{1}\ket{1_a}.
\eea
Thus, we have\\
\parbox{15cm}{
\begin{eqnarray*}
U_3\ket{u_0}\ket{k_a} &=& \sqrt{1-2\gamma^2}\ket{+}\ket{0_a} + \overline{\sqrt{-\gamma^2}}\ket{1}\ket{1_a} + \gamma\ket{2}\ket{0_a},\\
U_3\ket{u_1}\ket{k_a} &=& \sqrt{1-2\gamma^2}\ket{-}\ket{0_a} + \sqrt{-\gamma^2} \ket{1}\ket{1_a} + \gamma\ket{2}\ket{0_a}\\
U_3\ket{u_2}\ket{k_a} &=&\ket{2}\ket{0_a}.
\end{eqnarray*}}\hfill
\parbox{1cm}{\begin{eqnarray}\label{eq:transformation-3}\end{eqnarray}}\\
After the joint transformation, the quantum state we consider in discrimination is given by
\be\label{eq:mixed state-3}
\rho_\gamma = \sum_{i=0}^2p_iU_3(\out{u_i}{u_i}\ot\out{k_a}{k_a})U_3^{\dagger}.
\ee
By performing a von Neumann measurement on the auxiliary system by basis, $\{\out{0_a}{0_a}, \out{1_a}{1_a}\}$, the state in (\ref{eq:mixed state-3}) will collapse to either $\out{0_a}{0_a}$ or $\out{1_a}{1_a}$. If the system collapses to $\out{0_a}{0_a}$, we will discriminate the original state since those two states $\ket{u_0}, \ket{u_1}$ can be decided completely by the states $\ket{\pm}$ and the state $\ket{u_2}$ be decided uncertainly by the state $\ket{2}$ in (\ref{eq:transformation-3}). If the qubit collapses to $\out{1_a}{1_a}$, then we can only decide that the state is not $\ket{u_2}$.  when the qubit collapses to $\out{1_a}{1_a}$. Thus, we can design a mixed form of ambiguous and unambiguous discriminations as  follows:\\
\parbox{15cm}{
\begin{eqnarray*}
\pi_0 = \out{+}{+}\ot\out{0_a}{0_a}, & & \pi_1 = \out{-}{-}\ot\out{0_a}{0_a}, \quad \pi_2 = \out{2}{2}\ot\out{0_a}{0_a}\\
\textrm{and} & &
\pi_{fail} = \id_{\cH_2}\ot\out{1_a}{1_a},
\end{eqnarray*}}\hfill
\parbox{1cm}{\begin{eqnarray}\label{eq:measurement}\end{eqnarray}}\\
and the success probability of $\{\ket{u_i}\}_{i=0,1,2}$ can be discriminated is
\bea
P_{suc} = (1-2\gamma^2)(p_0+p_1) + p_2 = 1-2\gamma^2(1-p_2).
\eea

Moreover, we have

\begin{thrm}  Let $\cH_2 = \mathbb{C}^3$ and prepare three states $\{\ket{u_i}\}_{i=0,1,2}$ in $\cH_2$ with a priori probability distribution $p = (p_i)$, $\iinner{u_2}{u_0} = \iinner{u_2}{u_1} = \gamma$, $\iinner{u_0}{u_1} = 0$,  where $\gamma$ is a real number and $\gamma\neq 0$. If $p_2\geq\frac{1}{3}$, then
\bea
P_{suc} > P^{una}_{suc,max}.
\eea
\end{thrm}
\begin{proof} Following (\ref{eq:transformation-2}), we consider a unambiguous discrimination for those three states $\{\ket{u_i}: i=0,1,2\}$ with a priori probability distribution $p = \{p_i\}_i$ by coupling $\cH_2 = \mathbb{C}^3$ to $\cH_A$ by the joint unitary transformation $U_2$ as following:
\be\label{eq:transformation}
U_2\ket{u_i}\ket{k_a} = \sqrt{1-|\alpha_i|^2}\ket{i}\ket{0_a} + \alpha_i\ket{\phi_i}\ket{1_a},
\ee
where $\{\ket{\phi_i}, i=0,1,2\}\subseteq \cH_2$, and satisfy that
$\overline{\alpha_2}\alpha_0\iinner{\phi_2}{\phi_{0}} = \overline{\alpha_2}\alpha_1\iinner{\phi_2}{\phi_{1}} =
\gamma$ and $\iinner{\phi_0}{\phi_{1}} = 0$. Now, we decompose $\alpha_2\ket{\phi_2} = \alpha'_0\ket{\phi_0} + \alpha'_1\ket{\phi_1} + \beta\ket{\varphi}$, where $\overline{\alpha'_0}\alpha_0 = \overline{\alpha'_1}\alpha_1 = \gamma$ and $\iinner{\phi_1}{\varphi} = \iinner{\phi_2}{\varphi} = 0.$ Then, the success probability of unambiguous discrimination is given by
\bea
P^{una}_{suc} = 1 - p_0|\alpha_0|^2 - p_1|\alpha_1|^2 - p_2(|\alpha'_0|^2+|\alpha'_1|^2+|\beta|^2).
\eea
Note that the success probability of discrimination is the largest when $\beta = 0,$ thus, we find the optimal measurement. Therefore, we can rewrite (\ref{eq:transformation}) as
\bea
U_2\ket{u_0}\ket{k_a} &=& \sqrt{1-|\alpha_0|^2}\ket{0}\ket{0_a} + \alpha_0\ket{\phi_0}\ket{1_a},\\
U_2\ket{u_1}\ket{k_a} &=& \sqrt{1-|\alpha_1|^2}\ket{1}\ket{0_a} + \alpha_1\ket{\phi_1}\ket{1_a},\\
U_2\ket{u_2}\ket{k_a} &=& \sqrt{1-|\alpha'_0|^2-|\alpha'_1|^2}\ket{2}\ket{0_a} + \alpha'_0\ket{\phi_0}\ket{1_a} + \alpha'_1\ket{\phi_1}\ket{1_a}
\eea
where $\overline{\alpha'_0}\alpha_0 = \overline{\alpha'_1}\alpha_1 = \gamma.$ The success probability of unambiguous discrimination is given by
\be\label{eq:una. pro.}
P^{una}_{suc} = 1 - p_0|\alpha_0|^2 - p_1|\alpha_1|^2 - p_2(|\alpha'_0|^2+|\alpha'_1|^2).
\ee
Then, by $\overline{\alpha'_0}\alpha_0 = \overline{\alpha'_1}\alpha_1 = \gamma$ and $\max\{|\alpha_0|,|\alpha_1|,|\alpha'_0|,|\alpha'_1|\}\leq 1,$ we have that
\bea
P^{una}_{suc} < 1 - p_0\gamma^2 - p_1\gamma^2 - 2p_2\gamma^2 = 1 - \gamma^2(1 + p_2).
\eea
This showed that $P^{una}_{suc} < P_{suc}$ when $p_2 \geq \frac{1}{3}.$ The success probability (\ref{eq:una. pro.}) is applied in any unambiguous discrimination for the states $\{\ket{u_i}: i=0,1,2\}$, thus we have $P^{una}_{suc,max} < P_{suc}$ when $p_2 \geq \frac{1}{3}.$
\end{proof}

\begin{remark}\label{rem:separable form}
When $p_0 = p_1$, $\rho_\gamma$ is the state of separable form as follows
\be\label{eq:separable form}
\rho_\gamma = \{1 - \gamma^2(1-p_2)\}\rho_1^{\cH_2}\ot\out{0_a}{0_a} + \gamma^2(1-p_2)\out{1}{1}\ot\rho_2^{\cH_A},
\ee
where $\rho_1^{\cH_2}$ and $\rho_2^{\cH_A}$ are the density matrices of the principal system and the auxiliary system respectively,
\bea
\rho_1^{\cH_2} &=& \frac{1}{1 - (1-p_2)\gamma^2}\big\{\frac{1}{2}(1-p_2)(1-2\gamma^2)(\out{+}{+} + \out{-}{-}) + \big((1-p_2)\gamma^2+p_2\big)\out{2}{2}\\
&& + \frac{\sqrt{2}}{2}(1-p_2)\gamma\sqrt{(1-2\gamma^2)}(\out{0}{2}+\out{2}{0})\big\},\\
\rho_2^{\cH_A} &=& \frac{1}{(1-p_2)\gamma^2}\big\{(1-p_2)\gamma^2\out{1_a}{1_a} + \frac{\sqrt{2}}{2}(1-p_2)\sqrt{-\gamma^2(1-2\gamma^2)}\out{0_a}{1_a}\\
&& + \frac{\sqrt{2}}{2}(1-p_2)\overline{\sqrt{-\gamma^2(1-2\gamma^2)}}\out{1_a}{0_a}\big\}.
\eea
Thus, the discrimination of three states can be performed with the absence of entanglement. And, from (\ref{eq:separable form}) and the necessary and sufficient condition of zero discord in Ref. \cite{Dakic}, we have zero left quantum discord because that $[\rho_1^{\cH_2}, \out{1}{1}] = 0$. But, if $|\gamma| \neq \frac{1}{\sqrt{2}}$, the right discord is non-zero.
\end{remark}
\section{Generalization of the mixed form discrimination}
Next, we consider a general case, that is, let $\iinner{u_2}{u_0} = \iinner{u_2}{u_1} = \gamma,$ $\iinner{u_0}{u_1} = \alpha$, where $\gamma, \alpha$ be real numbers, and $\gamma\neq 0, 1; \alpha \neq 0, 1$.
Let us define
\bea
\ket{v_i} \equiv \ket{u_i}\ket{k_a}.
\eea
Taking two states $\ket{\psi_3},\ket{\psi_4}$ such that $\iinner{v_2}{\psi_3} = \iinner{v_2}{\psi_4} = 0$, and
\bea
\ket{v_0} &=& \sqrt{1-\gamma^2}\ket{\psi_3} + \gamma\ket{v_2},\\
\ket{v_1} &=& \sqrt{1-\gamma^2}\ket{\psi_4} + \gamma\ket{v_2}.
\eea
Note that $\iinner{v_0}{v_1} = \alpha = (1-\gamma^2)\iinner{\psi_3}{\psi_4} + \gamma^2$, we denote
\bea
c^2 = \iinner{\psi_3}{\psi_4} = \frac{\alpha-\gamma^2}{1-\gamma^2}.
\eea

Now, we couple $\cH_2 = \mathbb{C}^3$ to $\cH_A$ by a joint unitary transformation $U_4$ such that $U_4\ket{v_2} = \ket{2}\ket{0_a}$ and
\bea
U_4\ket{\psi_3} &=& \sqrt{1-|c|^2}\ket{+}\ket{0_a} + \overline{c}\ket{1}\ket{1_a},\\
U_4\ket{\psi_4} &=& \sqrt{1-|c|^2}\ket{-}\ket{0_a} + c\ket{1}\ket{1_a}.
\eea
Thus, we have
\bea
U_4\ket{u_0}\ket{k_a} &=& \sqrt{1-\gamma^2-|\alpha-\gamma^2|}\ket{+}\ket{0_a} + \overline{\sqrt{\alpha-\gamma^2}}\ket{1}\ket{1_a} + \gamma\ket{2}\ket{0_a},\\
U_4\ket{u_1}\ket{k_a} &=& \sqrt{1-\gamma^2-|\alpha-\gamma^2|}\ket{-}\ket{0_a} + \sqrt{\alpha-\gamma^2}\ket{1}\ket{1_a} + \gamma\ket{2}\ket{0_a},\\
U_4\ket{u_2}\ket{k_a} &=&\ket{2}\ket{0_a}.
\eea
After the joint transformation, the quantum state we consider in discrimination is given by
\be\label{eq:mixed state-4}
\rho_{\gamma,\alpha} = \sum_{i=0}^2p_iU_4(\out{u_i}{u_i}\ot\out{k_a}{k_a})U_4^{\dagger}.
\ee
Then, when $\alpha<\gamma^2$, by performing the von Neumann measurement such as (\ref{eq:measurement}), the success probability of $\{\ket{u_i}\}_{i=0,1,2}$ can be discriminated is
\bea
P_{suc, \alpha} = 1-(2\gamma^2-\alpha)(1-p_2),
\eea
when $\alpha\geq\gamma^2$, the success probability is
\be\label{eq:success probability}
P_{suc, \alpha} = 1-\alpha(1-p_2).
\ee

\begin{remark} When $\alpha<\gamma^2$ and $p_0 = p_1$, the quantum state (\ref{eq:mixed state-4}) is the state of separable form as follows
\be\label{eq:separable form-2}
\rho_{\gamma,\alpha} = \{1 - (1-p_2)(\gamma^2-\alpha)\}\rho_3^{\cH_2}\ot\out{0_a}{0_a} + (1-p_2)(\gamma^2-\alpha)\out{1}{1}\ot\rho_4^{\cH_A},
\ee
where $\rho_1^{\cH_2}$ and $\rho_2^{\cH_A}$ are the density matrices of the principal system and the auxiliary system respectively,
\bea
\rho_3^{\cH_2} &=& \frac{1}{1 - (1-p_2)(\gamma^2-\alpha)}\big\{\frac{1}{2}(1-p_2)(1+\alpha-2\gamma^2)(\out{+}{+} + \out{-}{-})\\
&& + \big((1-p_2)\gamma^2+p_2\big)\out{2}{2} + \frac{\sqrt{2}}{2}(1-p_2)\gamma\sqrt{(1+\alpha-2\gamma^2)}(\out{0}{2}+\out{2}{0})\big\},\\
\rho_4^{\cH_A} &=& \frac{1}{\gamma^2-\alpha}\big\{(\gamma^2-\alpha)\out{1_a}{1_a} + \frac{\sqrt{2}}{2}\sqrt{(\alpha-\gamma^2)(1+\alpha-2\gamma^2)}\out{0_a}{1_a}\\
&& + \frac{\sqrt{2}}{2}\overline{\sqrt{(\alpha-\gamma^2)(1+\alpha-2\gamma^2)}}\out{1_a}{0_a}\big\}.
\eea
Then, as Remark \ref{rem:separable form}, the discrimination of three states can be performed with the absence of entanglement. And, from (\ref{eq:separable form-2}) and the necessary and sufficient condition of zero discord in \cite{Dakic}, we have zero left quantum discord because that $[\rho_3^{\cH_2}, \out{1}{1}] = 0$.
But, the right discord is non-zero.
\end{remark}

\begin{thrm} Let $\iinner{u_i}{u_{j\neq i}} = \gamma \ \textrm{for}\  i,j = 0,1,2$, then
\bea
P_{suc, \gamma} \geq P_{suc,max}^{una}.
\eea
\end{thrm}

\begin{proof} Without lose of generality, we can assume $p_2 = \max\{p_i\}_{i=0,1,2}.$ By (\ref{eq:success probability}), we have $P_{suc, \gamma} = 1-\gamma (1-p_2) = 1 - \gamma (p_0 + p_1)$.

If the conditions (1) and (2) are satisfied in Example 1.2, then $P_{suc, \gamma} \geq P_{suc,max}^{una}$ is clear. If the condition (3) is satisfied in Example 1.2, note that $p_0 \geq p_0\gamma$ and $2\sqrt{p_1p_2} \geq p_1$, thus $P_{suc, \gamma} \geq P_{suc,max}^{una}$. If the condition (4) is satisfied in Example 1.2, note that the following inequalities:
\bea
p_0 \leq \sqrt{p_1p_2},\quad p_1 \leq \sqrt{p_1p_2} \quad \textmd{and} \quad \sqrt{p_0p_1} \leq \sqrt{p_0p_2}
\eea
where $p_0 \leq p_1 \leq p_2,$ we have that
\bea
P_{suc, \gamma} &=& 1-(p_0 + p_1)\gamma \geq 1 - 2\sqrt{p_1p_2}\gamma\\
&\geq& 1 - 2(\sqrt{p_1p_2} + \sqrt{p_0p_2} - \sqrt{p_0p_1})\gamma = P_{suc,max}^{una}.
\eea
\end{proof}

\begin{remark}
When $\alpha = \gamma^2$, it is possible to perform the above discrimination even without the auxiliary qubit system, because that the discrimination can be performed with the absence of both entanglement and quantum discord. This is also applied to following case:

Let $\iinner{u_i}{u_{j\neq i}} = \gamma_{ij}$ satisfy that $\overline{\gamma_{12}\gamma_{20}} = \gamma_{01}.$
Take two quantum states $\ket{\psi_0},\ket{\psi_1}$ such that $\iinner{u_2}{\psi_0} = \iinner{u_2}{\psi_1} = 0$ and
\bea
\ket{u_0} &=& \sqrt{1-|\gamma_{20}|^2}\ket{\psi_0} + \gamma_{20}\ket{u_2},\\
\ket{u_1} &=& \sqrt{1-|\gamma_{12}|^2}\ket{\psi_1} + \overline{\gamma_{12}}\ket{u_2}.
\eea
Thus, we have $\iinner{\psi_0}{\psi_1} = 0$ since $\overline{\gamma_{12}\gamma_{20}} = \gamma_{01}.$ Let us perform the measurment $\Pi_4 = \{\pi_i\}_i$ defined by
\bea
\pi_0 =\out{\psi_0}{\psi_0}, \ \  \pi_1 = \out{\psi_1}{\psi_1}\quad \textrm{and} \ \
\pi_2 = \out{u_2}{u_2}
\eea
on the state $\rho = \sum_{i=0}^3p_i\out{u_i}{u_i}$. Then, those two states $\ket{u_0}, \ket{u_1}$ can be decided completely when outcome is $i=0,1$, although the state $\ket{u_2}$ cannot be decided completely, but, we can decide it in following probability:
\bea
\frac{p_2\tr{\pi_2\out{u_2}{u_2}}}{p_0\tr{\pi_2\out{u_0}{u_0}} + p_1\tr{\pi_2\out{u_1}{u_1}} + p_2\tr{\pi_2\out{u_2}{u_2}}} = \frac{p_2}{p_0|\gamma_{20}|^2 + p_1|\gamma_{12}|^2 + p_2}.
\eea
\end{remark}

\subsection*{Acknowledgements}  This  project is supported by Research Fund, Kumoh National Institute of Technology, Korea.



\end{document}